\documentclass[12pt,preprint]{aastex}
\begin{document}

\title{Disk-Evaporation Fed Corona: Structure and Evaporation Feature with Magnetic Field}

\author{Lei Qian}
\affil{Astronomy Department, Peking University,
    Beijing, 100871, P.R. China}
\email{qianl@vega.bac.pku.edu.cn}

\author{B.F. Liu}
\affil{National Astronomical Observatories/Yunnan Observatory, Chinese
       Academy of Sciences, P.O. Box 110, Kunming 650011, China}
\and

\author{Xue-Bing Wu}
\affil{Astronomy Department, Peking University,
    Beijing, 100871, P.R. China}

\begin{abstract}
   The disk-corona evaporation model naturally interprets
   many observational phenomena in black hole X-ray binaries, such as the 
   truncation of an accretion disk and the spectral state transitions. On the 
   other hand,  
   magnetic field is known to play an important role in transporting angular 
   momentum and producing viscosity in accretion flows. 
   In this work, we explicitly take the magnetic field in the accretion disk 
   corona into account and numerically calculate the coronal structure on the 
   basis of our two-temperature evaporation code. We show 
   that the magnetic field influences the coronal structure by its contribution
   to the pressure, energy  and radiative cooling in the corona and by 
   decreasing the vertical heat conduction. We found that the maximal 
   evaporation rate keeps more or less 
   constant ($\sim 0.03$ Eddington rate) while the strength of magnetic fields 
   changes, but that the radius corresponding to the maximal evaporation rate 
   decreases with increasing magnetic field.  
   This predicts that the spectral state transition always occurs at a few 
   percent of Eddington accretion rate,  while the  inner edge 
   of thin disk 
   can be at $\sim 100 R_{\rm S} $ or even less in the hard state before the 
   transition to the soft state. These results alleviate the problem that 
   previous evaporation models predict too large a truncation radius, 
{and are in better agreement with
the observational results of several black hole X-ray binaries, though
discrepancies remain.}
\end{abstract}


\section{Introduction}

   X-ray binaries and Active Galactic Nuclei (AGN) are interesting objects in 
   astrophysics and have drawn a lot of 
   attentions since their discoveries. Accretion is widely accepted
   to be the main energy source of them
   (Shakura \& Sunyaev 1973, hereafter 
   SS73; Pringle 1981; Rees 1984). The thin disk model proposed by SS73
   not only successfully solves the energy problems in both X-ray binaries and 
   AGN, but also well 
   reproduces some key features of dwarf novae (Meyer \& Meyer-Hofmeister 
   1984). With the development of observing techniques in recent decades, 
   various types of broad band spectra have been observed in X-ray binaries 
   as well as in AGN.
   These spectra usually show a two-component feature, that is, a thermal 
   component together with 
   a non-thermal component  (see Remillard \& McClintock 2006 for a review of 
   X-ray binaries; see Mushotzky, Done \& Pounds 1993 for a review of AGN). 
   The thin disk alone, 
   which emits a multi-color blackbody spectrum, is difficult to 
   explain the
   observed broad band spectra. An additional hot accretion flow is required 
   to produce the high-frequency non-thermal spectrum. Therefore, a
   two-component model, consisting of an inner 
   advection dominated accretion flow  (ADAF, Narayan \& Yi, 1994), 
   or radiative inefficient accretion flow (RIAF, Yuan, Narayan \& Quataert 
   2003) and an outer Shakura-Sunyaev disk  (SSD, SS73), was proposed to explain
   the observational spectra and now has been widely adopted in studying 
   black hole binaries (e.g.Esin, 
   McClintock \& Narayan 1997; Quataert et al. 1999). However, there are still 
   some debates on
   what causes the transition between these two kinds of physically different
   accretion flows and how to determine the transition 
   properties, such as the critical accretion rate, the disk truncation radius,
   and so on.

   These problems have been extensively discussed in many previous studies 
   (Narayan \& Yi 1995; 
   Honma 1996; Meyer, Liu, \& Meyer-Hofmeister 2000b; Lu et al.  2004). 
   Among these models, only Meyer et al. (2000b) consider the detailed 
   vertical structure (a thin disk sandwiched by the corona) concerning the 
   transition between a SSD and a RIAF. It also  predicts a scale free 
   evaporation rate-radius ($\dot{m} -r$) relation, with which the disk 
   truncation radius is determined for a given accretion rate and is found to be
   consistent with observations (Liu et al. 1999; Liu \& Meyer-Hofmeister 2001).
   Further investigations show that the disk evaporation model can also 
   interpret the hysteresis of accretion rate between the hard-to-soft 
   transition 
   and the soft-to-hard transition, which has been observed in quite a 
   number of 
   low-mass X-ray binaries (Meyer-Hofmeister et al. 2005; Liu et al. 2005).
Recent studies (Liu, Meyer, \& Meyer-Hofmeister 2006; Meyer, Liu, \& Meyer-Hofmeister 2007) show  the disk evaporation model could explain the existence of an inner disk at intermediate states. 

   The original disk evaporation
   model was proposed by Meyer \& Meyer-Hofmeister (1994), where the magnetic 
   field is not included since the work aims at explaining the UV lag observed
   in dwarf novae. But magnetic field has long been realized to play an 
   important role in accretion disks (SS73). Recently, observations also 
   reveal the magnetic 
   nature of disk accretion onto black holes (Miller et al. 2006). Meyer \& 
   Meyer-Hofmeister  (2002) have included magnetic pressure in the disk 
   evaporation model in order to explain the different truncation radii of the 
   outer thin disks between the nuclei of elliptical galaxies and low 
   luminosity 
   AGN, which have similar Eddington-scaled accretion rate (normalized by the 
   Eddington accretion
   rate $\dot{M}_{\rm Edd}=L_{\rm Edd}/\eta c^2$, where $L_{\rm Edd}$ is 
   the Eddington luminosity, and $\eta$ is the efficiency of energy 
   conversion). 
   One temperature model(hereafter 1-T, which means that the electrons and ions 
   in the corona have the same temperature) was used in this previous study. 
   Besides the contribution of the magnetic pressure to the 
   total pressure, magnetic field can also affect the heat conduction. Meyer \&
   Meyer-Hofmeister  (2006) already took this effect into account together with
   the irradiation/Comptonization processes, and found that the variation 
   of heat conduction can change the $\dot{m}-r$ relation significantly.

   In this paper, we intend to investigate the influence of magnetic 
   field on both the pressure
   and heat conduction in detail with a more realistic two temperature model
   (hereafter 2-T, which means that the electrons and ions in the corona have
   different temperatures, Liu et al. 2002a). We present more detailed 
   numerical results and also give the fitting formula to the numerical results,
   which enables the comparison of theory and observations much easier.  
   In \S 2 we 
   briefly describe the model and  basic equations; in \S 3 we present the 
   numerical results; in \S 4 our results are compared with observations; in 
   \S 5 we discuss the model itself and present our conclusions in \S 6.

\section{Description of the model}
    Following previous works on disk evaporation (Meyer \& Meyer-Hofmeister 
    1994; Liu, Meyer \& Meyer-Hofmeister 1995; Meyer et al 2000a; 
    Liu et al. 2002a), we pay our main attentions 
    to the transition region between a SSD and a RIAF, which is characterized
    by a thin disk sandwiched by the corona flow. The viscous heat of ions is 
    partially transferred to electrons by collisions and conducted down by 
    electrons to the transition layer, which is cooler and denser.
    Part of the cool matter in the transition layer
    is heated and evaporated into the corona till an equilibrium density is 
    reached. The gas in the corona carries similar angular momentum as 
    that in the disk, and is continuously accreted to the central black 
    hole due to the viscous angular momentum transfer similar to that in 
    accretion disks. The drifting-in gas is steadily re-supplied by the 
    gas evaporating from the transition layer between the corona and the thin 
    disk, and an equilibrium state of the system can be reached.

    There are three factors crucial to the evaporation rate, i.e. the heating of
    the corona, the thermal conduction, and the radiative cooling in the 
    transition 
    layer. Any new physical process related to these three factors can
    influence the evaporation rate and eventually make the final 
    configuration of the accretion flow different (e.g. Liu, Meyer \& 
    Meyer-Hofmeister 2005).
    One important relevant factor is the magnetic field because it has 
    overall influence on accretion disks. Firstly, 
    it enhances the viscosity (Balbus \& Hawley 1991). Secondly, 
    magnetic field contributes to the total pressure. The viscous 
    heating changes by both the viscous coefficient and the pressure. 
    Thirdly, strong entangled magnetic field largely 
    modifies the heat conduction of the plasma (Tao 1995; Chandran 
    \& Cowley 1998; Narayan \& Medvedev 2001). Fourthly, in the presence of 
    magnetic field, there should be cyclotron-synchrotron radiation. In 
    addition, 
    magnetic dissipation/reconnection can be an additional heating mechanism to the corona.

    However, how magnetic field influences the viscosity parameter $\alpha$
    is still unclear. The 
    contribution could already be included if $\alpha$ is as large as 0.3. The
    dependence of evaporation rate on the viscosity
    parameter has been discussed in  Meyer-Hofmeister \& Meyer (2001).     
    Here we concentrate our investigations on the influences
    of magnetic pressure and heat conduction on the disk evaporation process. 
    The cyclotron-synchrotron radiation is optically thick in our corona and 
    could  only be important when the electron temperature is higher than 
    $10^9$K, 
    which is the case in the upper boundary layers. Compared with the cooling 
    from heat conduction, the cooling from optically-thick synchrotron 
    radiation in the 
    upper layers can be neglected as long as the electron temperature is not
    too 
    much higher than $10^9$K, which is indeed the case in our computations  
    (see also Meyer \& Meyer-Hofmeister 2002). Thus, 
    the cyclotron-synchrotron radiation will not be included in our 
    calculations. For simplicity,
    we assume a chaotic magnetic field, which provides an 
    isotropic magnetic pressure in the corona.

    The basic equations and boundary conditions we adopted in this work are 
    the same as those in Liu et al. (2002a), only with some minor changes. 
    The viscosity parameter (SS73) is set to
    $\alpha=0.3$. As having been shown in many
    previous works, the characteristics of disk evaporation are independent 
    on the mass of central black hole, the results are the same for systems  
    with stellar-mass and supermassive black holes.
    For clarity, we reproduce the basic equations here.
    
   The mass continuity equation is:
   \begin{equation}
    \frac{d}{dz}\left(\rho v_z\right) = \eta_m\frac{2}{r}\rho v_r - \frac
    {2z}{r^2+z^2}\rho v_z,
    \label{continuity}
   \end{equation}
   where $\rho$, $v_z$, $v_r$ are the gas density, the vertical 
   and radial component of the velocity, respectively. The factor $\eta_m$ 
   equals to 1 in the evaporation model (Meyer-Hofmeister \& Meyer 2003). 
   
   The z-component of the momentum equation is:
   \begin{equation}
    \rho v_z \frac{d v_z}{dz} = -\frac{dP}{dz} - \rho\frac
    {GMz}{(r^2+z^2)^{3/2}},
    \label{zmomentum}
   \end{equation}
   where the total pressure $P$ includes both the gas pressure $P_g$ and the
   magnetic pressure $P_m$.

   The equation of heat conduction is:
   \begin{equation}
    F_c=-\kappa T_e^{5/2}\frac{dT_e}{dz}.
    \label{conduction}
   \end{equation}
   where $\kappa$ is the heat conduction coefficient, and $T_e$ is the 
   temperature of electrons.

   The energy equation of ions is:
$$
    \frac{d}{dz}\left\{\rho_i v_z\left[\frac{v^2}{2}+\frac{\gamma}
    {\gamma-1} \frac{P_i}{\rho_i} -\frac{GM}{(r^2+z^2)^{1/2}}\right]\right\}
$$
$$
    =\frac{3}{2}\alpha P\Omega-q_{ie}+\eta_E\frac{2}{r}\rho_i v_r
    \left[\frac{v^2}{2}+\frac{\gamma}
    {\gamma-1} \frac{P_i}{\rho_i} -\frac{GM}{(r^2+z^2)^{1/2}}\right]
$$
   \begin{equation}
    -\frac{2z}{r^2+z^2}\left\{\rho_i v_z\left[\frac{v^2}{2}+\frac{\gamma}
    {\gamma-1} \frac{P_i}{\rho_i} -\frac{GM}{(r^2+z^2)^{1/2}}\right]\right\}.
    \label{ionenergy}
   \end{equation}
   where $v$ is the modulus of the velocity vector; $\rho_i$ and $P_i$ are 
   the contribution of ions to the density and pressure, respectively; 
   $\Omega$ is the angular velocity of the 
   gas in the corona; $q_{ie}$ is the heat transfer rate from ions to 
   electrons due to the Coulomb coupling (Liu et al. 2002a), which can be 
   expressed as

   \begin{equation}
    q_{ie}=\left(\frac{2}{\pi}\right)^{1/2}\frac{3}{2}\frac{m_e}{m_p}
    \ln\Lambda \sigma_T c n_e n_i \left(k T_i-k T_e\right)
    \frac{1+T_*^{1/2}} {T_*^{3/2}},
    \label{iecoupling}
   \end{equation}
   where $m_e$, $m_p$, $T_i$ and $T_e$ have their usual meaning; $\sigma_T$
   is the Thomson scattering cross section and $\ln \Lambda=20$ is the Coulomb 
   logarithm.
$
   T_*=(k T_e/m_e c^2)(1+(m_e/m_p)(T_i/T_e)),
$
   $\gamma$ is the ratio of specific heats, and $k$ is the Boltzmann 
   constant (rather than the heat conduction coefficient $\kappa$ ). The 
   factor $\eta_E$ equals to $\eta_m + 0.5$
   (Meyer-Hofmeister \& Meyer 2003).

   The combined energy equation of ions and electrons is:
$$
    \frac{d}{dz}\left\{\rho v_z\left[\frac{v^2}{2}+\frac{\gamma}
    {\gamma-1} \frac{P}{\rho} -\frac{GM}{(r^2+z^2)^{1/2}}\right]
    +F_c\right\}
$$
$$
    =\frac{3}{2}\alpha P\Omega-n_e n_i L(T_e)+\eta_E\frac{2}{r}\rho v_r
    \left[\frac{v^2}{2}+\frac{\gamma}
    {\gamma-1} \frac{P}{\rho} -\frac{GM}{(r^2+z^2)^{1/2}}\right]
$$
   \begin{equation}
    -\frac{2z}{r^2+z^2}\left\{\rho v_z\left[\frac{v^2}{2}+\frac{\gamma}
    {\gamma-1} \frac{P}{\rho} -\frac{GM}{(r^2+z^2)^{1/2}}\right]\right\},
    \label{energy}
   \end{equation}
    where $\Lambda(T)$ is the radiative cooling function for a low-density,
    optically thin gas of cosmic abundances in the temperature range of
    $10^4-10^8$K and is taken to be the Bremsstrahlung radiation function 
    for $T>10^8$K (Raymond, Cox \& Smith 1976).

    Note that the last terms in the right side of Eq.(\ref{continuity}), 
    Eq.(\ref{ionenergy}), and Eq.(\ref{energy})
    are the flaring terms (Meyer \& Meyer-Hofmeister 1994).
    The pressure $P$ in these equations
    is no longer the gas pressure but the total pressure, so the sound speed
    ($P/\rho$) has an additional factor $1+(1/\beta)$ compared with the 
    sound speed without the magnetic pressure ($P_g/\rho$)
    (where $\beta\equiv P_g/P_m$ 
    is the ratio of gas pressure to magnetic pressure). The ratio of specific 
    heats
    $\gamma$ also changes with $\beta$, with the relation $\gamma= (5\beta
    +8)/ (3\beta+6)$  (cf. Appendix A of Esin 1997, but note that the 
    definition of 
    $\beta$ is different there). 

    Both $\beta$ and $\kappa$ affect the lower boundary conditions 
    in the corona,
    which are ( Meyer-Hofmeister \& Meyer 2006)
   \begin{equation}
    T_i=T_e=10^{6.5}K,
    \label{lowT}
   \end{equation}

   \begin{equation}
    F_c=-2.73\times 10^6 P\lambda^{1/2}
    / (1+\frac{1}{\beta}), 
    \label{lowFc}
   \end{equation}
    where $\lambda\equiv \kappa/\kappa_{\rm Sp}$ is the fraction of the 
    standard value (Spitzer value $\kappa_{\rm Sp}=10^{-6} 
    g\  cm\  s^{-3}K^{-7/2}$) 
    of the heat conduction coefficient (Meyer-Hofmeister \& 
    Meyer 2006). We integrate Eq. (\ref{continuity}) to Eq.(\ref{energy})
    until the upper boundary conditions

   \begin{equation}
    v_z=c_s,
    \label{upvz}
   \end{equation}

   \begin{equation}
    F_c=0
    \label{upFc} 
   \end{equation}
    are fulfilled.

\section{Numerical results}

   We consider chaotic magnetic fields and parameterize it by $\beta\equiv P_g/P_m$ (where $P_m\equiv B^2/24\pi$).  
   So a smaller $\beta$ value means stronger magnetic 
   fields. Under an assumption of equipartition
   between the gas pressure and  magnetic pressure, we have $\beta=1$; 
   MHD simulation yields $\beta\approx 12$ (Sharma et al. 2006);
   the case without magnetic field corresponds to $\beta=\infty$. 
   Assuming different strength of magnetic pressure and heat 
   conduction, we can numerically solve the vertical structure of the corona at 
   different radii. Our main results are discussed as follows.  
 
\subsection{The vertical structure of the magnetized corona} 

\begin{figure}
   \centering
   \plottwo{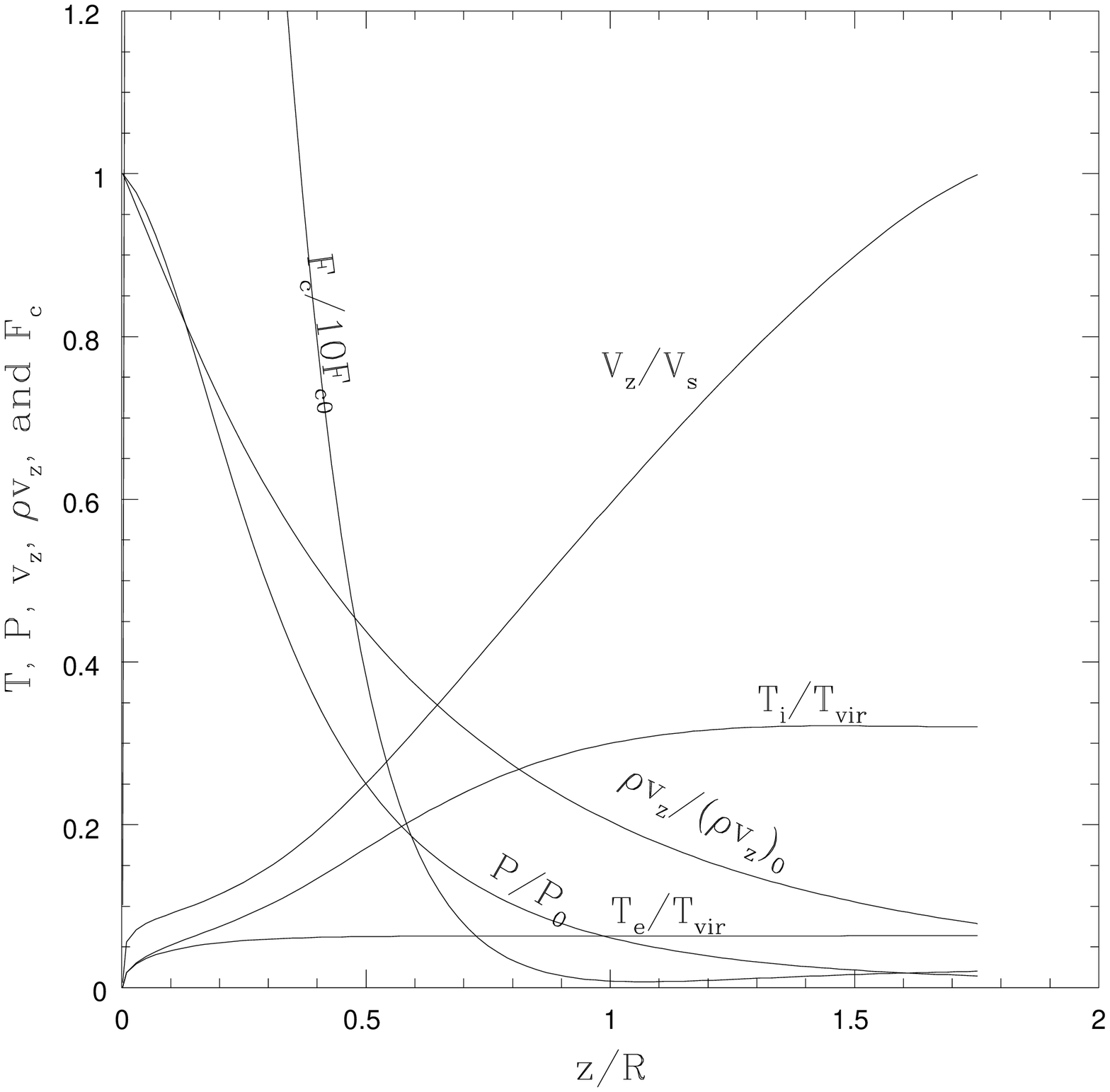}{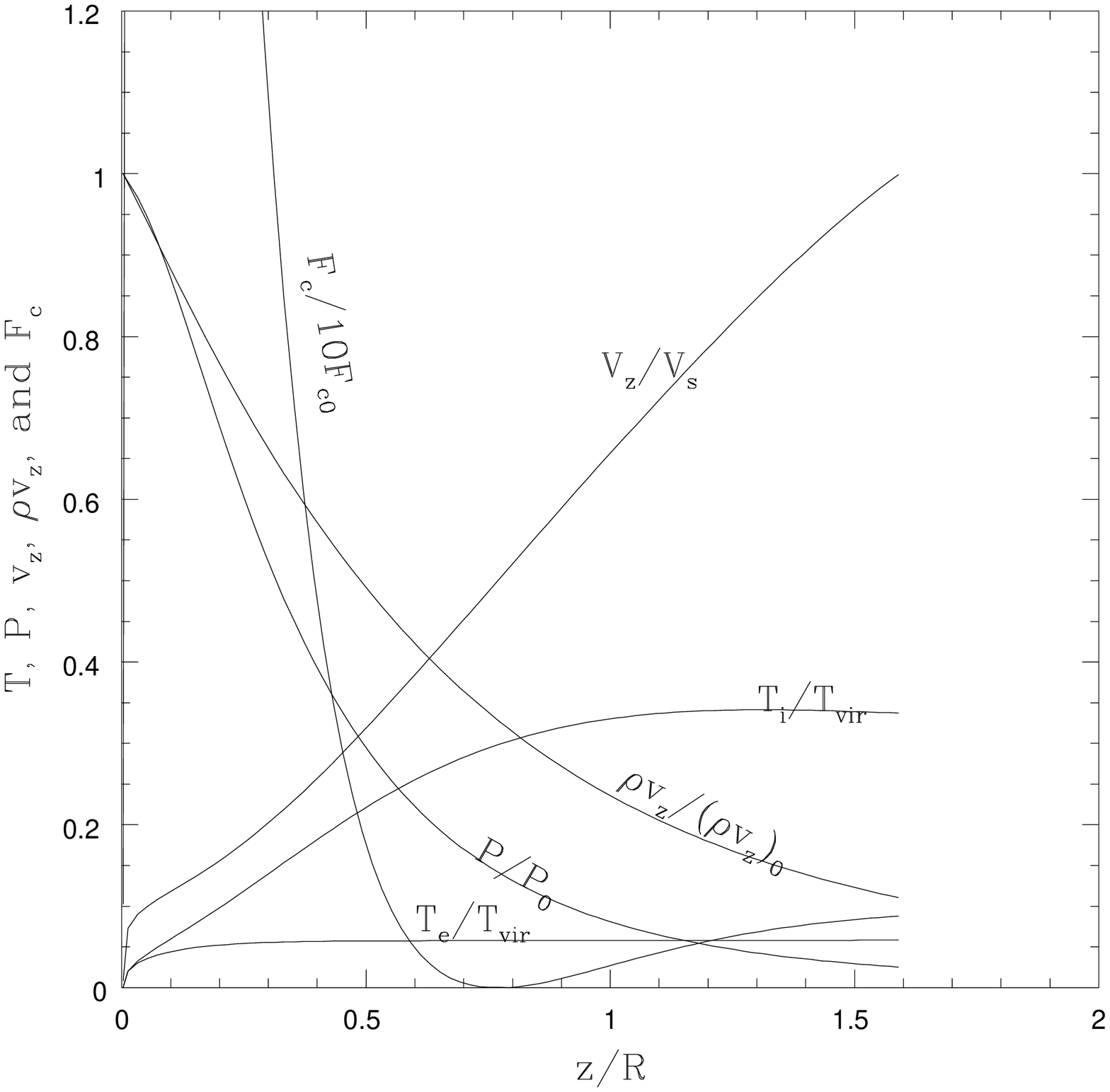}
      \caption{ 
                This figure shows the vertical structure of the corona  around 
                a black hole at a distance $R/R_{\rm S}=214$.
                $z$ is the height above the equatorial plane of the accretion 
                flow.
                $T_e$ and $T_i$ are electron temperature and ion temperature,
                respectively, while $T_{vir}$ is the virial temperature used
                for scale. $F_c$, $\rho v_z$, $P$, and $v_z$ are the heat
                flux, vertical mass flow rate per unit area, total pressure,
                and vertical velocity, respectively. $v_s$ is the sound
                speed used for scale. The quantities with subscript 0 are 
                the values at the lower boundary. 
                 Left: no magnetic field,
                $\beta=\infty$ and $\kappa=\kappa_{\rm Sp}$.
		Right: With magnetic field, $\beta=10/3$ and $\kappa=0.5\kappa_{\rm Sp}$}

         \label{structure}
   \end{figure}

Fig.~\ref{structure} shows the vertical structure of the corona 
   above a thin 
   disk at a radius $R=214R_{\rm S}$.  The left panel shows the structure without magnetic field, $\beta=\infty$  and $\kappa=\kappa_{\rm Sp}$. The right panel shows the influence of the magnetic field on the structure with   
  $\beta=10/3$  and $\kappa=0.5\kappa_{\rm Sp}$
Comparing the two cases, we find that the vertical profile of the coronal quantities do not change much when magnetic field is included, though the absolute values, for example,  the density and evaporation rate, do change.  This is a typical feature in the disk corona also shown in earlier works (e.g. Meyer et al 2000a), which indicates that the corona above a thin disk undergoes very steep changes in temperature, density, and can not be simply averaged in the vertical direction.

\subsection{The meaning of $\dot{m}-r$ curve}
   
   Qualitatively similar to previous results (Meyer et 
   al. 2000a; Liu et al. 2002a), our new calculations with the magnetic field show that the mass evaporating from the disk 
   to the corona 
   increases toward the central black hole, reaching a maximum at a few hundred 
   Schwarzschild radii and dropping very quickly near the black hole. The distribution of the evaporation rate along 
   the distance turns out to be independent on the black hole mass. We scale the
   evaporation rate by the Eddington accretion rate 
   $\dot{M}_{\rm Edd}\equiv L_{\rm Edd} /0.1c^2=1.39\times 10^{18}M/M_\odot$, 
   then $\dot{m}\equiv \dot{M}/\dot{M}_{\rm Edd}$, 
   and scale the distances by Schwarzschild radius $R_{\rm S}= 2GM/c^2$,
   then $r\equiv R/R_{\rm S}$. We will show  the dependence of $\dot m $  on $r$
   so that
   the relations $\dot m (r)$ can be tested by observations on some black hole 
   X-ray binaries.

   The $\dot{m}-r$ curve can predict the configuration of accretion flow 
   at different accretion rates. If the mass accretion rate of accretion disk
   is lower than the maximal evaporation rate, the disk will be depleted inside
   a radius where the accretion rate equals to the evaporation rate.
   This truncation radius is outside the radius of the maximal evaporation 
   rate. The accretion flow is then dominated by an inner pure corona, or more 
   exactly, the ADAF/RIAF. Such a configuration corresponds to the low/hard 
   spectral state in black hole X-ray binaries. However,it 
   is also possible that there is an inner disk which is separated from the 
   outer disk by a 'pure corona flow' if the accretion rate is close to the 
   maximal evaporation rate (Liu et al. 2006). This 
   configuration is thought to be related to the intermediate spectral state. 
   If the mass accretion rate exceeds the maximal evaporation rate,
   the disk cannot be depleted anywhere, so it can extend to the 
   marginal stable 
   radius (Meyer et al. 2000a). The radiation is dominantly from the standard 
   thin disk and the corona is suppressed by the Compton cooling. This situation
   corresponds to the high/soft spectral state.

   The change of mass accretion rate in the disk can also cause the transition 
   between different spectral states. As an example, if the mass accretion 
   rate in the disk increases from a value smaller than the maximal (critical)
   evaporation rate to a value larger than that, a hard-to-soft state 
   transition 
   would occur, and vise versa. So the maximal evaporation rate corresponds
   to a transition accretion rate (or a transition luminosity). This transition 
   accretion rate and the truncation radius of the disk in the low/hard, as
   shown in our predicted $\dot{m}-r$ curves,
   are the two parameters to be compared with the observational results. 

\subsection{The influence of magnetic field on $\dot m(r)$}

   We investigate the influence of magnetic pressure by 
   calculating the evaporation rate for a series of $\beta$ values.
   Our results based on both 1-T model and   
   2-T model are present in the $\dot{m}-r$ curves.

\begin{figure}
   \centering
   \plotone{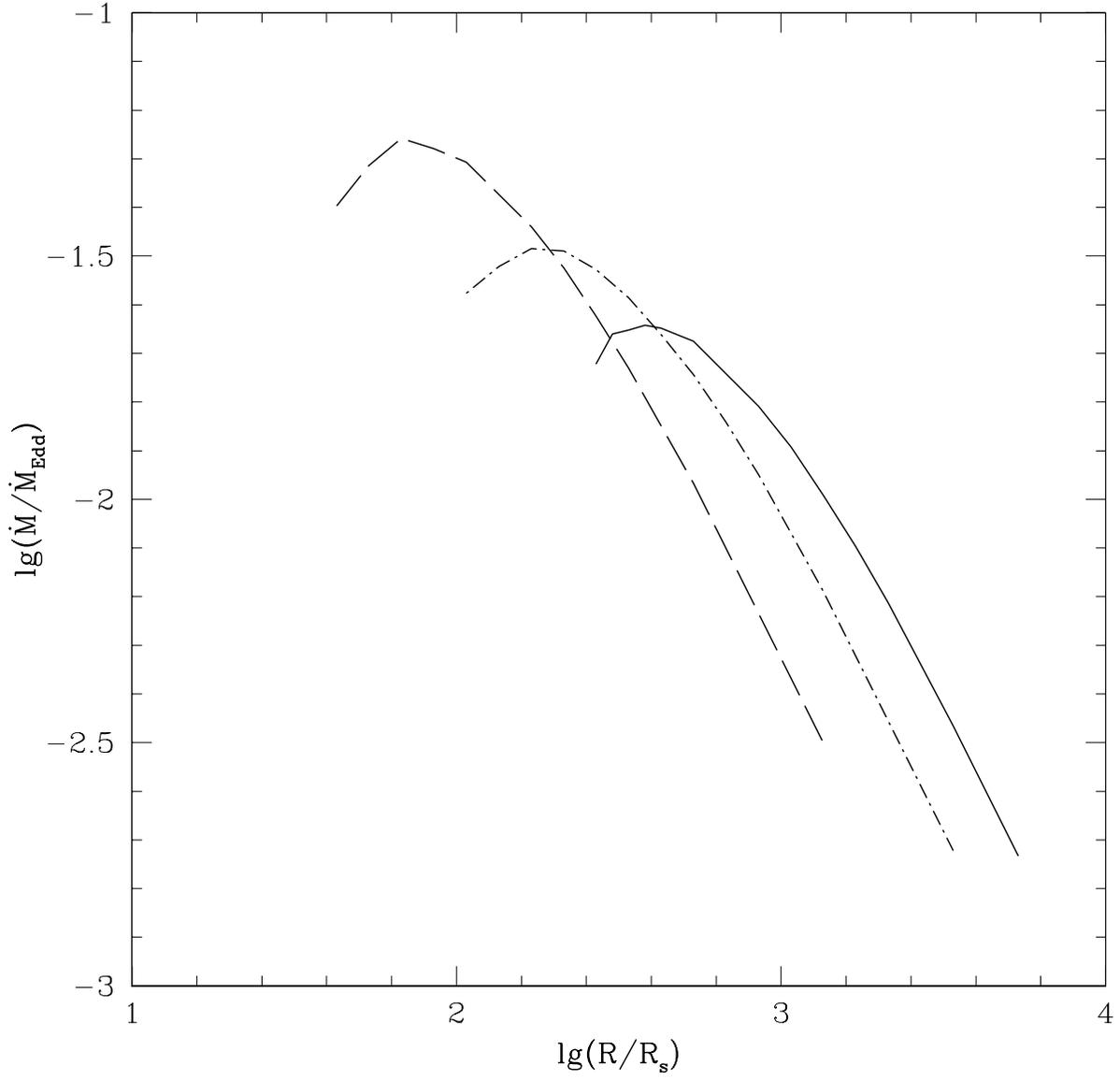}
      \caption{The $\dot{m}-r$ relation for different $\beta$ values  
               in the 1-T model. 
               The solid line, dash-dotted line, and long-dashed line 
               represent the cases of $\beta=\infty$, $\beta=5$, and $\beta=2$
               respectively.  The decease of $\beta$
               value corresponds to the increase of magnetic field strength.
              }
         \label{1T_beta}
   \end{figure}

   The $\dot{m}-r$ relation of 1-T model is shown in 
   Fig.~\ref{1T_beta}.
   The solid line, the dot-dashed line, and the long-dashed line represent 
   the cases of $\beta=\infty$ (
   no magnetic field),  $\beta=5$, and $\beta=2$ respectively. 
   We didn't calculate the curves with even smaller $\beta$, since magnetic 
   field would be too strong in those cases, invalidating the assumptions 
   of our current model. 
   The most obvious changes by the inclusion of magnetic field happen to  the 
   value 
   and location of the maximal evaporation rate. From Fig.~\ref{1T_beta} we see 
   that in the 1-T model, the  maximal evaporation rate increases with the 
   increase of magnetic field, while its location moves inward. This can 
   be understood as a consequence 
   of the enhanced viscous heating due to the contribution of magnetic pressure.
   The viscous heat cannot be efficiently consumed by radiative cooling until 
   a higher density is 
   reached in the inner region. There is also another influence lying in the 
   outer region (at large radii) of the corona. In contrast to the the maximal 
   evaporation rate,  the 
   evaporation rate at a fixed radius increases with the decrease of magnetic 
   field. This is because that the viscous heating is balanced by the advection 
   rather than by radiation in the outer region of the corona.  
   The viscous heating contributed by the 
   magnetic pressure is mainly transfered to enthalpy of gas which makes the gas
   hotter than the case without magnetic field and hence radiative cooling is 
   more efficient if the density is unchanged. This leads to a reduced 
   evaporation rate. These trends are consistent with
   the qualitative estimations given by Meyer et al. (2000a)
   (See their Eq. (56) and Eq. (57)).
{These evaporation features  are basically 
relying on enhanced dissipation rate produced by  the  enhanced magnetic pressure, 
which  might be true in other dissipation models as well, including those based on
reconnection and shocks in a strongly magnetized corona.}

   {Such a distribution of evaporation rate along distance, as shown in  Fig.~\ref{1T_beta}, }
   implies that at a given accretion 
   rate the truncation of disk occurs at smaller radius with the involving of 
   magnetic field and the disk can extend to a  smaller radius before the 
   system transits to the soft state.   

\begin{figure}
   \centering
   \plotone{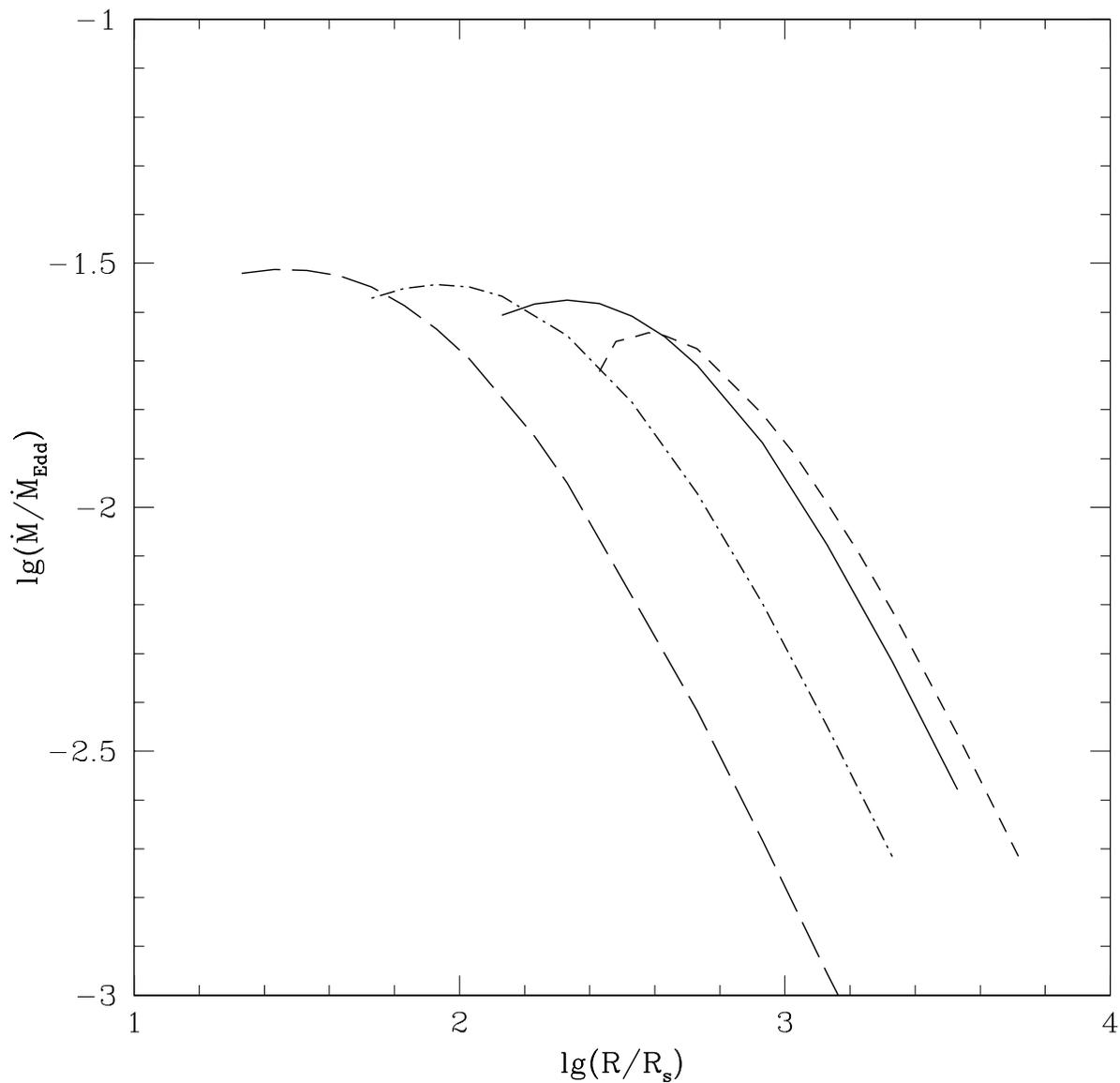}
      \caption{The $\dot{m}-r$ relation for different $\beta$ values 
               in the 2-T model. 
               The short-dashed line represents the case of $\beta=\infty$ in 
               the 1-T model, which is plotted here 
               for comparison. Other lines represent the curves in the 
               2-T model. The solid line, dot-dashed line, and long-dashed line
               represents the cases of $\beta=\infty$, $\beta=5$, and $\beta=2$
               respectively.
              }
         \label{2T_beta}
   \end{figure}

   In the 2-T model we take into account the decoupling of electrons and ions 
   probably occurring in the inner region.  When electrons and ions are 
   decoupled, the electron heating by collision with ions is inefficient. 
   A large 
   fraction of heat is stored in ions and the heat to evaporate the gas in 
   the disk is thus limited. Therefore, a smaller evaporation rate is 
   expected in 
   the 2-T scenario than that in the 1-T case, especially in the inner region.
   Our numerical results are shown in  Fig.~\ref{2T_beta}.  
   The lines other than the short-dashed line represent the results calculated 
   for the 2-T model, with the same notations as in Fig.~\ref{1T_beta} for the 
   1-T model. As expected, the 1-T results and 2-T results are similar in the 
   outer region. However, there are clear differences in the inner 
   region where ions and 
   electrons are not well coupled. Because the 2-T model usually describes 
   the real physics in the inner hot accretion flow, we will concentrate our 
   discussion on the 2-T model.
   Fig.~\ref{2T_beta} shows that the curve for evaporation rate 
   shifts inwards with the increase of magnetic field,  while the 
   maximal evaporation rates keep more or less the same, around 2-3\% of the
   Eddington rate. This indicates that the involve of magnetic field results in
   a smaller truncation radius of the thin disk. The stronger the magnetic 
   field is, the smaller 
   radius the disk is truncated at. However, our result implies that the 
   transition from the hard state to the soft 
   state occurs at almost the same accretion rate, no matter the magnetic 
   field is included or not. Such a feature may interpret why some objects 
   truncate at smaller radius than  that predicted by previous models 
   (see section 4). 
   To show more detailed influence of magnetic field  we list the maximum 
   evaporation rates and the corresponding radii for different $\beta$ values 
   in  
   the upper part of Table ~\ref{mdot-r}. The data can be linearly fitted by 
   $\log \dot m_{\rm max}=0.143/
   \beta-1.579$ and $\log r_{\rm max}=-1.750/\beta+2.299$.

\subsection{The influence of heat conduction $\dot m(r)$}

   Theoretical works have shown that the chaotic magnetic field would suppress 
   the heat conduction in the plasma (Tao 1995; Chandran \& Cowley 1998; 
   Narayan \& Medvedev 2001). Recent work reports that the coefficient of 
   heat conduction can be reduced to one fifth of the Spitzer value 
   (Narayan \& Medvedev 2001).
   We calculate the $\dot{m}-r$ relation for different $\kappa$ with both
   the 2-T model and 1-T models. The 2-T results do not include the 
   irradiation/Comptonization effects since we are more interested in the 
   influence caused by the heat conduction. {The hard radiations from the inner corona/ADAF can heat up electrons (irradiation) in most of the corona region wheras cool electrons (Comptonization) in the innermost region, thereby lead to an outward shift of the evaporation curve $\dot m(r)$ (Liu et al. 2005), counteracting partially the effect of reduced heat conduction.
For details of} the combined effect of heat 
   conduction and irradiation/Comptonization one can refer to the work of 
   Meyer-Hofmeister \& Meyer (2006).

\begin{figure}
   \centering
   \plotone{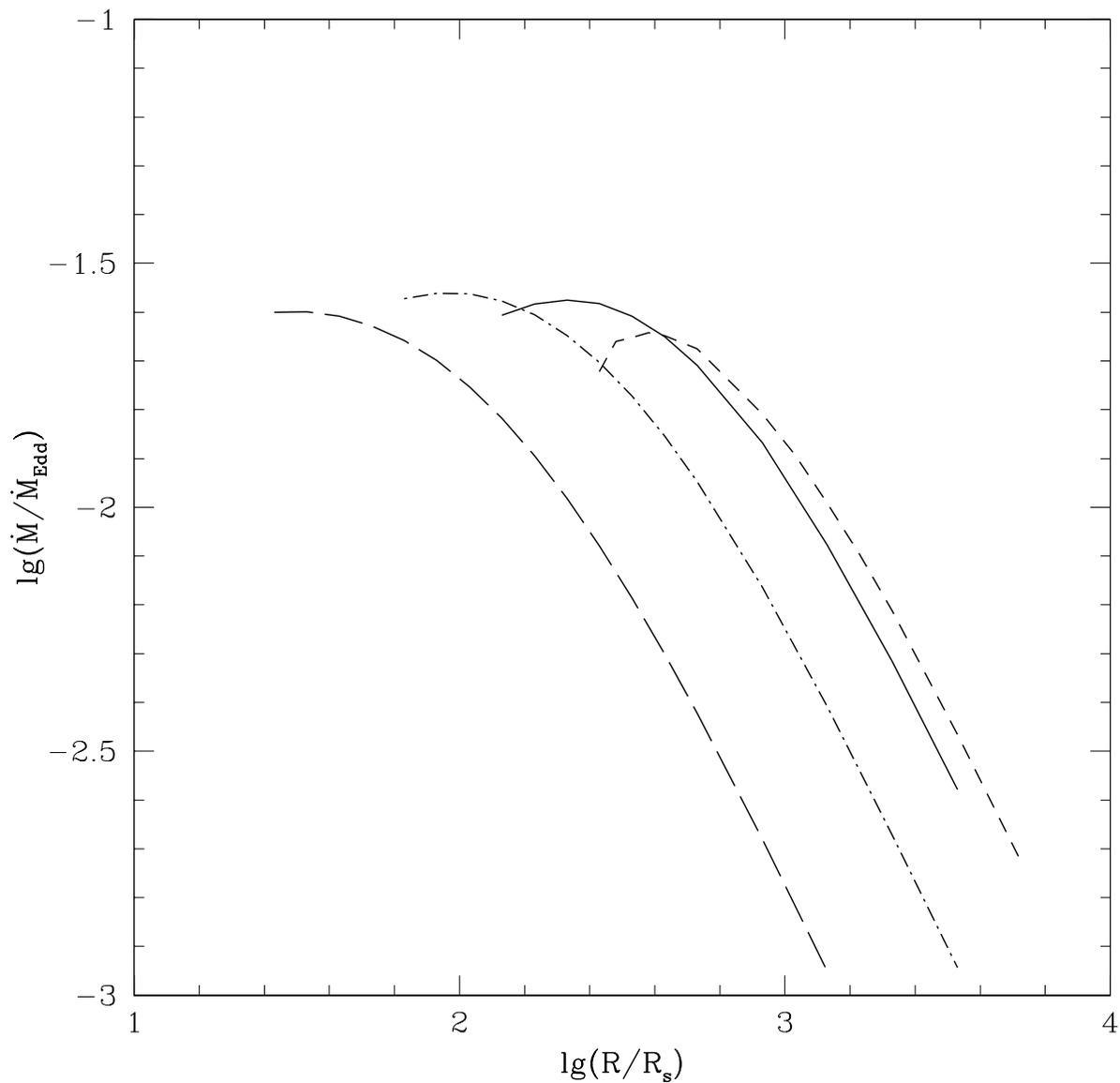}
      \caption{The $\dot{m}-r$ relation for different $\kappa$ values 
               in the 2-T model.
               The short-dashed line is for $\kappa=\kappa_{\rm Sp}$  of the 
               1-T model and is plotted here for comparison. 
               The solid line, dot-dashed line, and long-dashed line
               represent the cases of $\kappa=\kappa_{\rm Sp}$, 
               $\kappa=0.5 \kappa_{\rm Sp}$, and $\kappa=0.2\kappa_{\rm Sp}$,
               respectively.
              }
         \label{2T_kappa}
   \end{figure}

   Fig.~\ref{2T_kappa} shows the dependence of the $\dot{m}-r$ relation on 
   $\kappa$ based on the 2-T model.
   The solid line represents the standard case with $\kappa=\kappa_{\rm Sp}
   $, . The dot-dashed line and the long-dashed line correspond
   to $\kappa=0.5 \kappa_{\rm Sp}$, and  
   $\kappa=0.2 \kappa_{\rm Sp}$, respectively . For comparison, a  curve with 
   $\kappa= \kappa_{\rm Sp}$ in the 1-T model is also shown as a short-dashed 
   line.  Fig.~\ref{2T_kappa} looks very similar to Fig.~\ref{2T_beta}. The 
   location of the maximal evaporation rate moves inward with the decrease of 
   $\kappa$, but the maximal evaporation rate is insensitive to  
   $\kappa$. Detailed values of our calculation and the corresponding linearly
   fitting results are listed in the lower half of Table ~\ref{mdot-r}.
   Comparing the fitting results for both the cases of magnetic pressure and 
   heat conduction, 
   we find that the influence of magnetic field is much stronger than that of 
   the heat
   conduction. Note that the chaotic magnetic field  tends to decrease the heat 
   conduction by deflecting the motion of electrons and the inclusion of 
   magnetic 
   field would  greatly reduce the disk truncation radius through both $\beta$
   and $\kappa$. We'll show a composite result in the next subsection.  

\begin{figure}
   \centering
   \plotone{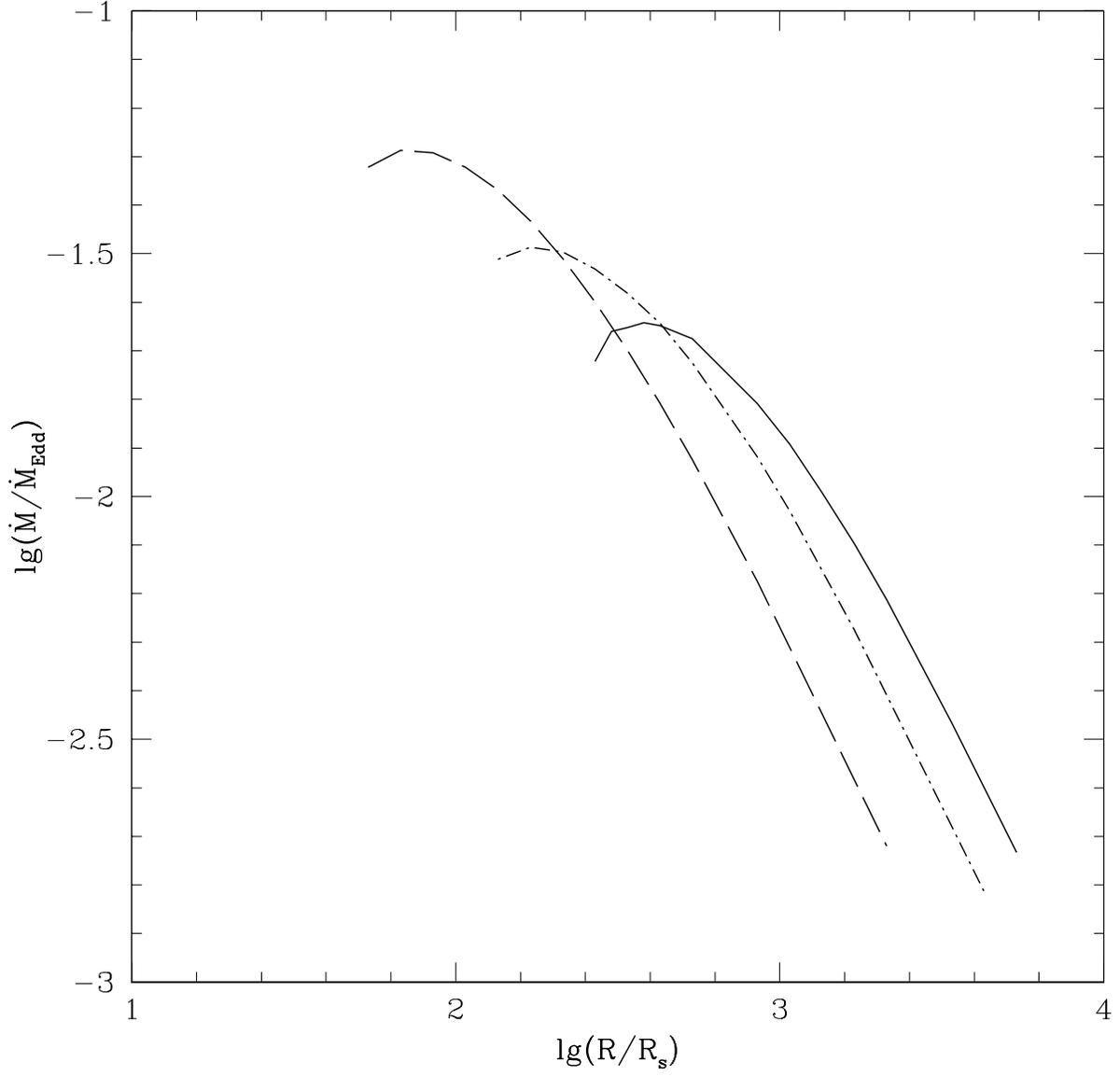}
      \caption{The $\dot{m}-r$ relation for different $\kappa$ values in 
               the 1-T model, 
               The solid line, dot-dashed line, and long-dashed line
               represent the cases of $\kappa=\kappa_{\rm Sp}$.
               $\kappa=0.5 \kappa_{\rm Sp}$, and $\kappa=0.2\kappa_{\rm Sp}$,
               respectively.
              }
         \label{1T_kappa}
   \end{figure}

   The 1-T results with different $\kappa$ values are plotted in  
   Fig.~\ref{1T_kappa}, which shows the similar tendency as the influence of 
   $\beta$. Due to the decoupling of electrons and ions in the inner 
   region, the 1-T results are less meaningful than the 2-T results.

\subsection{The combined effect of $\beta$ and $\kappa$}

\begin{figure}
   \centering
   \plotone{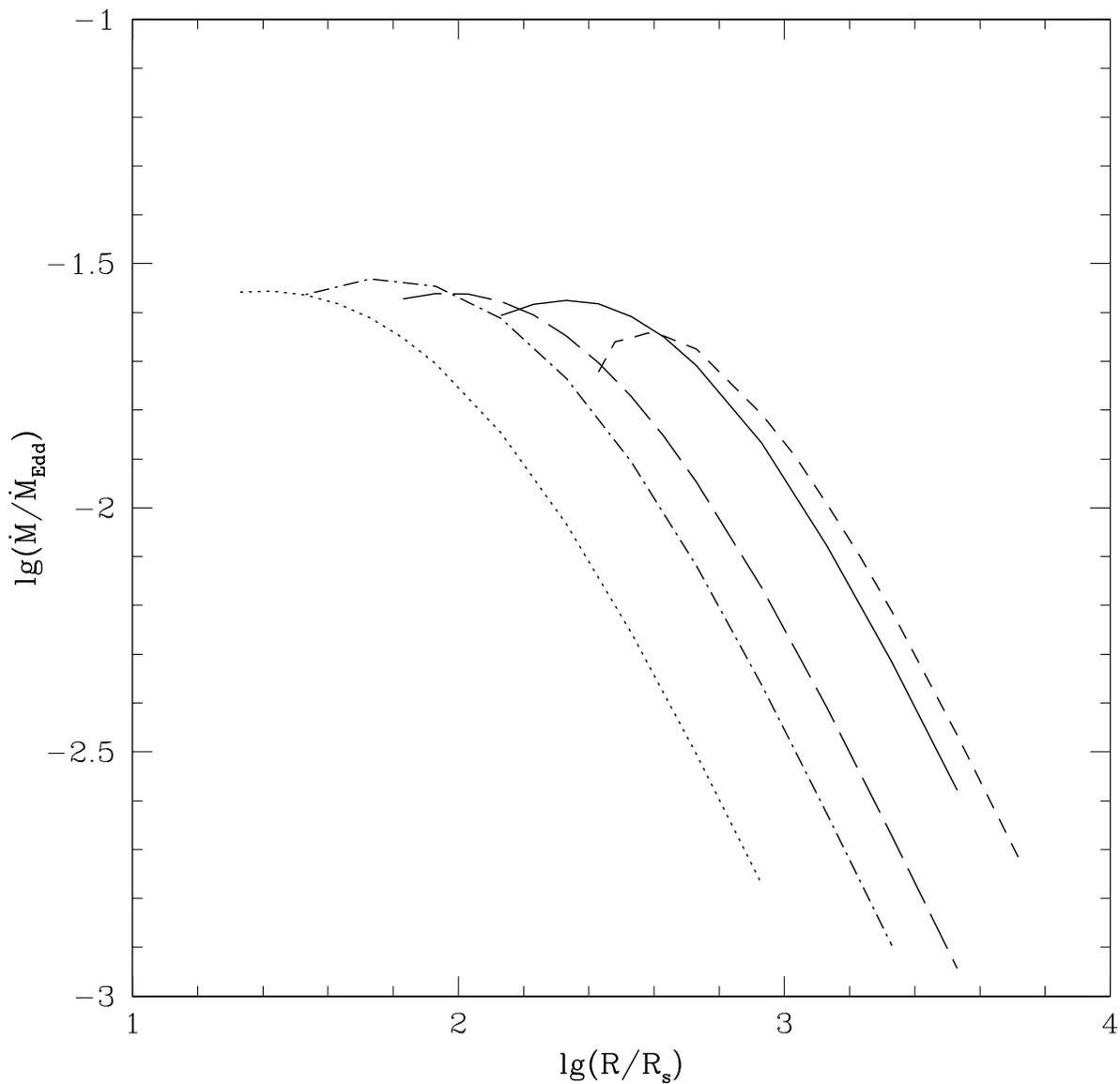}
      \caption{The composite influence of $\beta$ and $\kappa$ in the 2-T model,
               The solid line, long-dashed line, dash-dotted line 
               and dotted line
               represent the cases of ($1/\beta=0$, $\kappa=\kappa_{\rm Sp}$), 
               ($1/\beta=0$, $\kappa=0.5 \kappa_{\rm Sp}$),
               ($1/\beta=0.3$, $\kappa=\kappa_{\rm Sp}$), and 
               ($1/\beta=0.3$, $\kappa=0.5 \kappa_{\rm Sp}$)
               , respectively. The short-dashed 
               line represents the result for $1/\beta=0$ and  
               $\kappa=\kappa_{\rm Sp}$ in the 1-T model, which is plotted here
               for comparison.
              }
         \label{beta_kappa}
   \end{figure}

  Qualitatively the presence of chaotic magnetic field can reduce the 
  efficiency of heat 
  conduction by electrons. But how $\kappa$ changes with $\beta$ remains 
  unclear. As an example, we take $1/\beta=0.3$  
  and  $\kappa=0.5 \kappa_{\rm Sp}$ to calculate the 
  evaporation rates at different radii. The results are shown in 
  Fig.~\ref{beta_kappa} (the dotted line). One can see that the 
  magnetic pressure and the heat conduction play similar roles to the shift of 
  $\dot m(r)$. That is, the increase of magnetic field or the decrease of heat 
  conduction shifts the evaporation curve $\dot m(r)$ inwards. Their combined 
  effects are the superposition of the shifts caused separately by the 
  magnetic pressure and the heat conduction. However, the maximal evaporation 
  rate increases slightly with magnetic pressure and even more slightly with 
  heat conduction. Since the magnetic field tends to reduce the efficiency of 
  heat conduction, the two effects on the maximal evaporation rate counteracts
  each other in the case of the inclusion of magnetic field, thereby keeping 
  the maximal rate almost constant. 
  The maximal evaporation rate influenced by the
  combination  of $\beta$ and $\kappa$ can be derived roughly from the 
  combined linear fitting to the two sets of data together with the data from 
  the case of ($\beta=0.3$, $\kappa/\kappa_{\rm Sp}=0.5$), namely, 

\begin{equation}
\log \dot m_{\rm max}=0.101/\beta+0.023\kappa/\kappa_{\rm Sp}-1.589,
\label{mdot-k-b}
\end{equation}

\begin{equation}
\log \dot m_{\rm max}=-0.272\log\beta'+0.037\log(\kappa/\kappa_{\rm Sp})-1.566,
\label{mdot-logk-logb}
\end{equation}
  where $\dot m_{\rm max}=\dot M_{\rm max}/\dot{M}_{\rm Edd}$ and
  $\beta'=\beta/(\beta+1)$.

  The location of the maximal evaporation rate can be obtained in the same way, 
  namely,

 \begin{equation}
\log r_{\rm max}=-1.935/\beta+0.899\kappa/\kappa_{\rm Sp}+1.466,
\label{r-k-b}
 \end{equation}

 \begin{equation}
\log r_{\rm max}=5.158\log\beta'+1.170\log(\kappa/\kappa_{\rm Sp})+2.333,
\label{r-logk-logb}
 \end{equation}
  where $r_{\rm max}=R_{\rm max}/R_{\rm S}$.

  Eqs.(\ref{mdot-k-b}) to (\ref{r-logk-logb}) show more clearly than 
  Fig.~\ref{beta_kappa}
  that the maximal evaporation rate hardly changes with the inclusion of 
  magnetic 
  field and heat conduction; while the truncation radius depends 
  strongly on both the magnetic pressure and heat conduction. This implies, if 
  there exists magnetic field in the disk, the truncation of the disk could 
  occur 
  at much smaller radii, while the hard-to-soft state transition should occur at
  more or less the same accretion rate, $\dot m\approx 0.03$. For instance, in 
  the case of $1/\beta=0.3$, $\kappa=0.5\kappa_{\rm Sp}$ (shown by the dotted 
  line in Fig.~\ref{beta_kappa}), the hard/soft transition occurs at 
  $\dot m=0.028$ and the disk can extend down to $\sim 27 R_{\rm S}$ 
  before the transition. Such a transition radius is much smaller than 
  the prediction of 
  $\sim 209 R_{\rm S}$ without  involving the magnetic field.

  If the dependence of $\kappa$ on magnetic field is known, we can more 
  accurately
  determine the maximal evaporation rate and the corresponding radius, and 
  make the comparison of  
  the theoretical predictions with the accretion rate and inner disk 
  radius determined by observations during the hard-to-soft state transition.

\section{Comparison of the model predictions with Observations}

   Our numerical results can be compared with observations.
   Since the results predicted by the disk evaporation model are mass
   scale-free, they can be used in various systems with different masses of 
   central black holes.

\subsection{Luminosities at spectral state transitions in X-ray binaries}   

   The maximal evaporation rate in the $\dot m-r$ curve predicts a critical 
   accretion rate at which a 
   transition occurs between the hard and soft spectral states.  When the 
   accretion rate in the disk is lower than the critical one, the evaporation 
   depletes the 
   disk and the accretion flow is eventually replaced  by an ADAF/RIAF. 
   When the accretion rate is higher than the critical value, the standard 
   thin disk can extend down to the last stable 
   orbit and thus dominates the accretion flows. From Fig.~\ref{2T_beta} and 
   Fig.~\ref{2T_kappa}, one can 
   see that this transition accretion rate is around $0.03\dot{M}_{\rm Edd}$ 
   on the basis of the 2-T model 
   (the 1-T model gives $0.02\dot{M}_{\rm Edd}$, see 
   also Meyer et al. 2000a), which is weakly dependent on the strength of 
   magnetic pressure and heat conduction. This 
   quantity can be compared with the transition luminosities in black hole 
   X-ray binaries
   if the efficiency of energy conversion $\eta$ is similar.

   The observed transition luminosities (scaled by the Eddington luminosity) 
   in some black hole X-ray binaries
   are listed in Table ~\ref{tluminosity}. One can see that different objects 
   have transition luminosities ranging from 1\% to 15\%. The average value 
   of the transition luminosities is 0.036. However, when we talk about 
   the transition 
   luminosities, we must take special cautions. Firstly, the luminosities of 
   different sources are observed at different energy band and may not 
   represent the bolometric
   luminosities. Secondly, these observation values of transition luminosities
   suffer from the uncertainties of some parameters, such as the masses of the 
   central black holes and
   the source distances(Gierli\'nski \& Newton 2006). If these 
   dimensionless transition luminosities can represent the 
   corresponding dimensionless mass accretion rates, we can estimate their 
   average value as 0.036. This value predicted by the disk 
   evaporation model ($\sim 0.03$) is well consistent 
   with the observational result.

\subsection{Truncation radius and the corresponding accretion rate at hard 
            states}
   In the hard state of black hole X-ray binaries, 
   the accretion rate in the disk is lower than the maximal evaporation 
   rate, and the thin disk will be truncated at a radius where the accretion
   rate in the disk equals to the evaporation rate. So the 
   $ \dot{m}-r $ relation
   can be tested by the truncation radius of the disk and the
   corresponding accretion rate estimated from the observational data.

   For such a purpose, we collect some data from black hole X-ray binaries (see 
   Table ~\ref{truncation}).  These data are based on the spectral fitting with 
   the ADAF + disk model (Narayan, Barret \& McClintock 1997). The accretion 
   rates in Table ~\ref{truncation} are the Eddington-scaled values 
   ($\dot m=0.1 \dot M c^2/L_{\rm Edd}$), which are converted from the Eddington
   ratios ($L/L_{\rm Edd}$)(Zdziarski et al. 2004),  or from the rates scaled by
   the critical accretion rate ($ \dot{m}\equiv \dot{M}c^2/L_{Edd} $,  Wilms et
   al. 1999), or from $\dot M$ (in the unit of solar mass per year)
   (Poutanen et al. 1997).
   The data listed in Table ~\ref{truncation} can then be compared
   with the theoretical results from our model.

\begin{figure}
   \centering
   \plotone{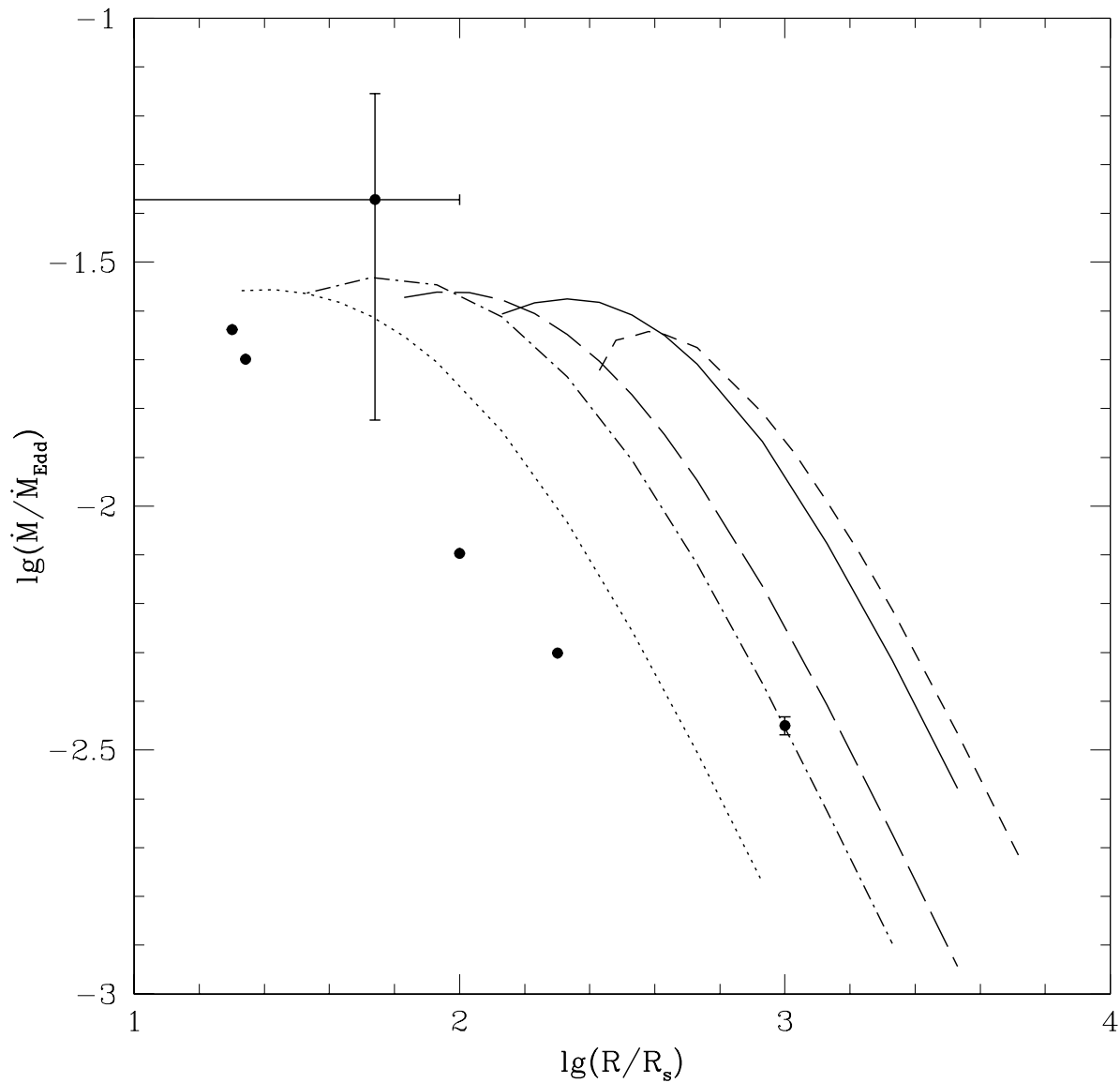}
      \caption{A comparison of the theoretical results and observations, 
               The solid line represents the case of $1/\beta=0$
               and $\kappa=\kappa_{\rm Sp}$.
               The long-dashed line represents the case of $\kappa=0.5
               \kappa_{\rm Sp}$ and $1/\beta=0$.
               The dot-dashed line corresponds to the case of $1/\beta=0.3$
               and $\kappa=\kappa_{\rm Sp}$.
               The dotted line is for $1/\beta=0.3$ and $\kappa=0.5
               \kappa_{\rm Sp}$.
               The filled dots represent the data listed in 
               Table ~\ref{truncation}.
              }
         \label{2T_observe}
   \end{figure}

   Fig.~\ref{2T_observe} shows the observational data together with our model 
   predictions for different strength of magnetic pressure and heat conduction. 
   One can see that the enhanced magnetic pressure and reduced heat conduction 
   bring the 
   model predictions much closer to the observations.  We expect that 
   theoretical results with stronger 
   magnetic field or further reduced heat conduction or both will predict more 
   consistent results with observations. Here we don't give such an example
   not only because we don't know the accurate strength of magnetic field in 
   individual objects, but also due to the large uncertainties in the 
   observational
   data. Caution should be taken here that there are several sets of data given
   in the original papers, we just listed the best fitted ones here. 
   We also noticed that  {in modeling the low-state spectrum of XTE J1118+480 by an inner ADAF and an outer disk (Esin et al. 2001), the ratio   
   $P_g/(P_g+P_m)$ in the ADAF is set to 0.97}, which corresponds to $\beta=32$ is this work.

\section{Discussion}
\subsection{Comparison with earlier estimations}    
   Our 1-T calculations confirm the results estimated by Meyer et al. (2000a)
   , namely,
   \begin{equation}
    \left(\frac{\dot{M}}{\dot{M}_{\rm Edd}}\right)  \propto  \frac{\alpha^3}{
    \kappa^{1/2}}\left(1+\frac{1}{\beta}\right)^{5/2},
    \label{maxm}
   \end{equation}
   \begin{equation}
    \left(\frac{R}{R_{\rm S}}\right)  \propto  \frac{\kappa}{
    \alpha^2}\left(1+\frac{1}{\beta}\right)^{-4}.\\
    \label{maxr}
   \end{equation}
    However, our results from the 2-T model are different from that of the 
    1-T model. The maximal evaporation rate
    remains almost the same with the decrease of both $\beta$ and $\kappa$, 
    though the location of maximal evaporation rate moves to smaller 
    radii in a similar way to that in the 1-T model. In the outer region, 
    both the 2-T model and the 1-T 
    model give similar results, since the accretion flow is close to 
    a 1-T flow. However, in the inner region, ions and electrons are decoupled.
    The accretion energy
    is mainly stored in the ions and the temperature of ions is higher than 
    that of electrons. Therefore, the accretion flow is a 2-T flow.
    There is little heat conducted by the electrons from corona to the
    disk, and thus the matter evaporation is inefficient compared to the 
    1-T case.

    \subsection{Cooling by cyclo-synchrotron radiation}

    With the inclusion of magnetic field, one potential cooling mechanism
    is the cyclo-synchrotron radiation, which is negligible in the lower layers
    (Meyer \& Meyer-Hofmeister 2002). In the highest part of the corona,
    it may influence the temperature (one can refer to the discussion of
    synchrotron radiation in Narayan \& Yi (1995) and Mahadevan (1997)).  
    However, we expect little influence from  cyclo-synchrotron radiation on the main body
    of the coronal action.

\subsection{Heating by magnetic field dissipation}
    Besides the influence on some physical processes such as heat conduction, 
    the magnetic field may play an important role in the
    formation of corona (Galeev, Rosner \& Vaiana 1979). 
    Disk magnetic fields rising into the corona contain energy that 
    originates in the disk accretion but will be dissipated (e.g. by 
    reconnection and shocks) in the corona. We take this very roughly into 
    account by scaling the viscous dissipation in our modeling to the total 
    instead to the gas pressure and thus including parts which are related to 
    the addition of disk produced magnetic flux that raise the magnetic 
    pressure, i.e. lead to a lower beta in the corona.

\subsection{Comparison with MHD simulation}

   Detailed investigations on the formation of a magnetized corona from MHD
   simulation (e.g. Miller \& Stone 2000; Hawley \& Balbus 2002; 
   Machida, Hayashi, \& Matsumoto 2000) show that a strongly magnetized
   corona  can form above an initially weakly magnetized disk. 
   Miller \& Stone (2000) showed that 
   magnetic field is amplified within 2 disk scale height (H=0.01R) and the
   energy is mostly dissipated in 3 to 5 scale height, thereby heats up the
   corona. Machida et al. (2000) also showed that a low-$\beta$ corona in the 
   form of "patch corona" or active coronal region. 
   {More recent work (Hirose, Krolik, \& Stone 2006), which includes radiation
   transport, shows that magnetic dominance happens
 deeper than the photosphere, where the medium is still relatively cool.}
   Here we consider a  
   slab corona in a large vertical extent where the structure is vertically 
   stratified, in contrast to an isothermal torus in the MHD simulation.  
   More importantly, the vertical thermal conduction is taken into account in 
   our model. This leads to efficient mass evaporation from the disk to the 
   corona,  feeding a corona to higher mass density than that in the coronal 
   envelope shown in MHD simulations.  Therefore, we expect relatively higher 
   $\beta$ value in our case.

\section{Conclusion}
    We investigate the influence of coronal magnetic fields 
    on the structure of an accretion disk corona 
    sustained by dissipative energy release and thermal coupling between 
    corona and disk. Numerical calculations show that the relation between mass 
    evaporation rate and radius ($\dot m$ - r curve) systematically shifts to 
    smaller radii with both an increase of magnetic pressure (decreasing 
    $\beta$) and a decrease of heat conductivity (decreasing $\kappa$). 
    The location of maximal evaporation rate
    lies between $30R_{\rm S}$ and $200R_{\rm S}$ when $\beta$ ranges from 2 to 
    $\infty$ or $\kappa$ from $\kappa_{\rm Sp}$ to $0.2\kappa_{\rm Sp}$.
    However, the 
    maximal evaporation rate remains almost the same ($\sim 0.03 \dot{M}_{\rm 
    Edd}$, with energy conversion efficiency $\eta=0.1$)
    If the disk truncation and state transition are indeed 
    caused by an evaporation process, the transition luminosity predicted by 
    our disk evaporation model is 
    $L_{tr}=0.03 L_{\rm Edd}$ before transition to the soft state.
    The inner edge of the outer thin disk
    can be several ten to several hundred Schwarzschild radii, depending on 
    the strength of magnetic field and its effect on the heat conduction.
    This alleviates the problem that the previous evaporation models predict 
    too large a disk truncation radius before the transition from hard to soft 
    state.
    {Our predictions with the
inclusion of the coronal field are found to be in better agreement with
the observational results of several black hole X-ray binaries, though
discrepancies remain.}

\acknowledgments
       We thank F. Meyer and E. Meyer-Hofmeister for detailed discussions on the computational results and for their reading through the manuscript and comments on it. We also thank F.K. Liu for helpful discussions.  Q.L. wishes to thank the hospitality 
       of High Energy Astronomy Group at YNAO, which makes this work possible. 
       This work is partially supported  by the National Natural Science 
       Foundation of China (Grants-10533050, 10473001, 10525313), the 
       BaiRenJiHua program of the Chinese Academy of Sciences, and the RFDP 
       (Grant 20050001026).

\appendix

\clearpage

\begin{table}
\caption{Influences of magnetic pressure and heat conduction on the maximal 
         evaporation rate and its radius, followed by the best linear fitting
         results}
\label{luminosity}      
\centering                          

 \begin{tabular}{c c c c}\hline\hline
  $ 1/\beta $ ($\kappa/\kappa_{\rm Sp}=1.0$) 
& $\log\beta'=\log(\beta/(\beta+1))$
& $\log(R_{\rm max}/R_{\rm S})$  
& $\log(\dot{M}_{\rm max}/\dot{M}_{\rm Edd})$  \\
\hline

  0.0 &  0.000 & 2.320 & -1.575 \\
  0.1 & -0.041 & 2.110 & -1.579 \\
  0.2 & -0.079 & 1.920 & -1.544 \\
  0.3 & -0.114 & 1.780 & -1.530 \\
  0.4 & -0.146 & 1.615 & -1.520 \\
  0.5 & -0.176 & 1.420 & -1.513 \\

\hline
\multicolumn{4}{l}{$\log(R_{\rm max}/R_{\rm S})=-1.750/\beta+2.299$ }\\
\multicolumn{4}{l}{$\log(\dot{M}_{\rm max}/ \dot{M}_{\rm Edd})=0.143/
  \beta-1.579$ }\\
\multicolumn{4}{l}{$\log(R_{\rm max}/R_{\rm S})=4.971\log\beta'+2.321$  }\\
\multicolumn{4}{l}{$\log(\dot{M}_{\rm max}/ \dot{M}_{\rm Edd})=-0.408
   \log\beta'-1.581$  }\\
\hline\hline
  $ \kappa/\kappa_{\rm Sp} $ ($1/\beta=0.0$) 
& $\log(\kappa/\kappa_{\rm Sp})$
& $\log(R_{\rm max}/R_{\rm S})$ 
& $\log(\dot{M}_{\rm max}/ \dot{M}_{\rm Edd})$  \\\hline

  0.2 & -0.699 & 1.540 & -1.600 \\
  0.5 & -0.301 & 1.990 & -1.561 \\
  0.6 & -0.222 & 2.070 & -1.561 \\
  0.7 & -0.155 & 2.160 & -1.563 \\
  0.8 & -0.097 & 2.200 & -1.567 \\
  0.9 & -0.046 & 2.290 & -1.571 \\
  1.0 &  0.000 & 2.320 & -1.575 \\

\hline
\multicolumn{4}{l}{$\log(R_{\rm max}/R_{\rm S})=0.956\kappa/\kappa_{\rm Sp}
                   +1.440$  }\\
\multicolumn{4}{l}{$\log(\dot{M}_{\rm max}/ \dot{M}_{\rm Edd})=0.024\kappa/
 \kappa_{\rm Sp}-1.588$  } \\
\multicolumn{4}{l}{$\log(R_{\rm max}/R_{\rm S})=1.124
  \log(\kappa/\kappa_{\rm Sp})+2.325 $  }\\
\multicolumn{4}{l}{$\log(\dot{M}_{\rm max}/ \dot{M}_{\rm Edd})=0.040
  \log(\kappa/\kappa_{\rm Sp})-1.563 $  } \\
\hline
 \end{tabular}
    \label{mdot-r}
\end{table}

\clearpage

\begin{table}
\caption{Transition luminosity in black hole X-ray binaries}  
\label{tluminosity}      
\centering                          

 \begin{tabular}{c c c}\hline\hline
 Source & $L_{trans}/L_E$ & Ref. \\\hline
 Nova Mus 91 & 0.031 & 1  \\
 XTE J 1550-564 & 0.034 & 1  \\
 GS 2000+251 & 0.0069 & 1 \\
 Cyg X-1 & 0.028 & 1 \\
 GRO J 1655-40 & 0.0095 & 1 \\
 LMC X-3 & 0.014 & 1 \\
 XTE J 1550-564 & 0.03 & 2\\ 
 XTE J 1650-500 & 0.02 & 2\\
 XTE J 2012+381 & 0.02 & 2\\
 4U 1543-47 & 0.07 & 2\\
 GX 339-4 & 0.15 or 0.05 & 2\\
 GX 339-4 & 0.06 or 0.02 & 2\\
 GX 339-4 & 0.14 & 3\\\hline
  $\sum^n_{i=1}L_i/n$ & 0.036 & \\\hline
 \end{tabular}
\begin{list}{}{}

\item Reference: 1. From Maccarone (2003). Note: the luminosities of different  
         sources are at different energy bands. 2. From Gierli\'nski \& Newton 
         (2006). Note: the luminosity is in the 1.5-12 keV band. 3. From 
         Zdziarski et al. (2004).
\end{list}

\end{table}

\clearpage

\begin{table*}[t]
\caption{Accretion rates and truncation radii of disk determined from 
         observations}
\label{truncation}
\begin{tabular} {llll}\hline
Source & $R_{in}/R_{\rm S}$ & $\dot{m}$  $^{\mathrm{a}}$ & Reference 
 \\\hline
 GX 339-4 & 200 & 0.005   & Wilms et al. 1999
 \\
 GX 339-4 & 100 & 0.008   & Wilms et al. 1999
 \\
 GX 339-4 & $10\sim 100$ & $0.015\sim 0.07$  & 
 Zdziarski et al. 2004 \\
 GRO J1655-40 & 1000 & $0.0034\sim 0.0037$ 
 & Hameury et al. 1997  \\
 XTE J1118+480 & 55 & 0.02 
 & Esin et al. 2001  \\
 Cyg X-1 & 20 & 0.023 
 & Poutanen et al. 1997 \\
 \hline
\end{tabular}
\begin{list}{}{}
\item[$^{\mathrm{a}}$] These accretion rates are scaled by $\dot{M}_{\rm Edd}
                       \equiv L_{\rm Edd}/\eta c^2$ with $\eta=0.1$, namely 
                       $\dot{m}=\dot{M}/\dot{M}_{\rm Edd}$.
\end{list}
\end{table*}

\clearpage

\clearpage

\clearpage

\clearpage

\clearpage

\clearpage

\clearpage

\end{document}